\documentclass[aps,pra,showpacs,twocolumn,superscriptaddress]{revtex4-1}
\usepackage{bm,color,amsmath,amssymb,mathrsfs,latexsym,graphicx,psfrag,float}

\usepackage{ifthen}
\newboolean{insert_titles}
\setboolean{insert_titles}{false} 
\newcommand{\opt}[1]
{
  \ifthenelse{\boolean{insert_titles}}{#1}{}
}

\newcommand{\bra}[1]{\langle #1|}
\newcommand{\ket}[1]{|#1\rangle}

\newcommand{\changed}[1]{{\color{black}#1}}
 
\usepackage{color}
\newcommand{\etal}{\emph{et al.}}

\begin{document}

\title{Effects of Smooth Boundaries on Topological Edge Modes in Optical Lattices}

\author{Michael Buchhold}
\author{Daniel Cocks}
\author{Walter Hofstetter}
\affiliation{Institut f\"ur Theoretische Physik, Johann Wolfgang Goethe-Universit\"at, 60438 Frankfurt/Main, Germany}

\begin{abstract}
Since the experimental realization of synthetic gauge fields for neutral atoms, 
the simulation of topologically non-trivial phases of matter with ultracold atoms 
has become a major focus of cold atom experiments. However, several obvious 
differences exist between cold atom and solid state systems, for instance the 
small size of the atomic cloud and the smooth confining potential. In this 
article we show that sharp boundaries are not required to realize quantum Hall 
or quantum spin Hall physics in optical lattices and, on the contrary, that edge 
states which belong to a smooth confinement exhibit additional interesting 
properties, such as spatially resolved splitting and merging of bulk bands and 
the emergence of robust auxiliary states in bulk gaps to preserve the 
topological quantum numbers. In addition, we numerically validate that these 
states are robust against disorder. Finally, we analyze possible detection 
methods, with a focus on Bragg spectroscopy, to demonstrate that the edge states 
can be detected and that Bragg spectroscopy can reveal how topological edge 
states are connected to the different bulk bands.
\end{abstract}


\maketitle
\section{Introduction}
Ultracold atoms in optical lattices provide a unique experimental setup for 
studying properties of solid state systems in a very clean and well controlled 
fashion \cite{Bloch2008, Cooper2008}. Particularly interesting in this context is the experimental 
implementation of artificial gauge fields for neutral atoms 
\cite{Bloch2008, Lin2009, Lin2009b, Lewenstein2007}\changed{\cite{Cooper2012}}, simulating for instance 
time-reversal symmetry breaking magnetic fields \cite{Jaksch2003, Mueller2004, 
Sorensen2004, Spielman2009, Gerbier2010}\changed{\cite{JimenezGarcia2012}} or a coupling of the atom's internal 
spin degree to its angular momentum \cite{Dalibard2011,Lin2011,Goldman2011,Lim2010}. The realization of these effects will 
open a path for precise simulations of a large class of topologically non-trivial 
systems such as quantum Hall (QH) or quantum spin Hall (QSH) phases.  Creating 
topological states of matter with cold atoms is particularly attractive because 
of the precise control of physical parameters such as the hopping amplitude 
and interaction strength, allowing the possibility to observe strongly 
interacting topological phases in lattice experiments. However the 
implementation of artificial gauge fields for neutral atoms is only one 
experimental challenge in simulating topological phases in optical 
lattices \cite{Ruseckas2005,Osterloh2005,Aidelsburger2011, Struck2012}. Experiments must overcome the difficulties provided by the finite size 
of the lattice and the soft boundary of the system, caused by a trapping 
potential that is smoothly varying in space.  Finite size leads to a finite 
overlap of \changed{spatially separated counterpropagating edge states} and therefore to possible 
backscattering processes, decreasing the robustness of the edge states against 
external perturbations \cite{Stanescu2009, Stanescu2010}.  While this is not a very serious restriction for 
optical lattice potentials, which are relatively pure, the effects of the soft boundary 
of the optical lattice system may significantly change the properties of the 
edge states characterizing topological insulators in finite systems.  
Whereas recent publications identify the soft boundaries as an unwanted 
restriction or propose how to avoid them by implementing artificial sharp 
boundaries to their system \cite{Goldman2010a}, we demonstrate in this article that soft boundaries 
will lead to interesting additional features, either not present or at least not visible in 
systems with sharp boundaries. For this purpose, we investigate different trap 
shapes and geometries, which are realizable in optical lattices and discuss 
their specific influence on the cold atom system.

This article is organized in the following way.
First, in Sec.~\ref{sec2}, we present the theoretical model under consideration, 
a QH Hamiltonian in the tight-binding approximation for spin polarized fermions 
confined in an additional trapping potential. In Sec.~\ref{sec3}, we present our 
results for the stripe geometry, discussing in detail the properties of the edge 
states in systems with a hard wall boundary, a harmonic trap and a quartic 
trapping potential. In Sec.~\ref{sec4}, we study the shape of the edge states in 
a completely trapped system and investigate the suitability of several detection 
methods as tools to probe the system experimentally, including Bragg 
spectroscopy.  Finally, in Sec.~\ref{sec5}, we provide some conclusions.

\section{The model}\label{sec2}
The model we consider is similar to the \changed{ones proposed in \cite{Jaksch2003, Gerbier2010}, 
experimentally realizing time-reversal symmetry breaking topological insulators with 
ultracold atomic gases.  This model describes a two-dimensional (2D) system of 
spin-polarized fermionic atoms subjected to a square optical lattice, experiencing an 
artificial Abelian gauge field $\bm{A}$ that induces an artificial uniform 
magnetic field perpendicular to the lattice, $\bm{B}= B\bm{e}_z$, which is similar to 
the celebrated Hofstadter Hamiltonian \cite{Hofstadter1976} on the lattice}. In our system,
the gauge field $\bm{A}$ enters the 
first-quantized Hamiltonian of the system in form of the minimal coupling 
$\bm{p} \rightarrow \bm{p}-\frac{e}{c}\bm{A}$, which leads to the Hamiltonian:
\changed{\begin{equation}
\label{Hfirst}
H=(\mathbf{p}-\frac{e}{c}\mathbf{A})^2/2m+W(\mathbf{x})+V(\mathbf{x}).
\end{equation}} Hamiltonian \eqref{Hfirst} contains \changed{the optical lattice potential $W$ and a spatially dependent scalar 
potential $V$} which allows for the inhomogeneity of the lattice, caused by the 
finite width of the laser beams creating the lattice or additional external 
potentials such as a harmonic trap or an artificial hard wall boundary. For the 
moment we leave the detailed shape of $V$ arbitrary, and only assume that the 
non-local matrix elements of $V$ are negligible (i.e. $\bra l V\ket 
m=\delta_{l,m}\bra l V\ket l $, where $\bra l$ is the Wannier state at lattice 
site $l$), which is reasonable in our case since the potential is either varying 
slowly compared to the lattice spacing $a$, or is a step function. 

The second quantized form of Hamiltonian \eqref{Hfirst} 
in the tight binding approximation then reads
\begin{equation}
\label{Hamiltonian1}
H=-t\sum_{l,m} c^{\dagger}_l e^{i2\pi\phi_{l,m}}c^{\phantom{\dagger}}_m+\sum_l 
V_l c^{\dagger}_l c^{\phantom{\dagger}}_l.
\end{equation}
The operator $c_l^{\dagger}$ here denotes the fermionic creation operator at 
lattice site $l$, with its respective annihilation operator 
$c_l^{\phantom{\dagger}}$. The first term is the well known nearest neighbor 
(NN) hopping with amplitude $t$, and is complex due to the Peierls phases 
$2\pi\phi_{l,m}$ that are a result of the gauge field \cite{Hofstadter1976}.  
The second term corresponds to the inhomogeneity $V$ with the 
local matrix elements $V_{l}\equiv \bra l 
V \ket l$.  The phases 
$\phi_{l,m}=\frac{1}{2\pi}\int_l^m\mathbf A \cdot d\mathbf l$ are not uniquely 
defined by the magnetic field and depend on the gauge chosen. In this paper, we 
choose the common Landau gauge $\mathbf A=(0,Bx,0)$, which leads to 
$\phi_{l,m}=\alpha\cdot x_l (\delta_{y_l,y_m+1} - \delta_{y_l,y_m-1})$, where 
$x_l$ and $y_l$ are the coordinates of lattice site $l$ with lattice spacing 
$a=1$ and $\alpha=\frac{\Phi}{\Phi_0}$ represents the flux per plaquette in 
units of the magnetic flux quantum, $\Phi_0=h/e$.  Setting $e=\hbar=1$, we 
obtain $\alpha=\frac{B}{2\pi}$ for the square lattice.  Throughout the rest of 
this article we choose the hopping $t$ as the natural energy unit of our system.

In the following sections we will restrict our analysis to the case where 
$\alpha=1/6$ or $\alpha=2/5$ respectively. Our results for these two cases can 
easily be generalized to other cases where $\alpha=p/q$, with $p,q\in 
\mathbb{N}$, and where topological edge states are predicted 
\cite{Hatsugai1993}. 

\changed{The experimental realization of a similar model was proposed in \cite{Goldman2010a},
where the authors consider a spinful fermionic system subjected to an artificial gauge field
that simulates a magnetic field of the form $\bm{B}= B\sigma \bm{e}_z$, where $\sigma=\pm 1$ is
the spin quantum number. This model preserves time-reversal invariance and therefore allows for the
realization of QSH phases in optical lattices. Because of the time-reversal 
symmetry, our analysis also applies to this model when spin remains a good 
quantum number, and we will mention the corresponding QSH phases throughout the 
text.}
So far, we have not accounted for a Zeeman splitting due to an external magnetic 
field, a spin-orbit coupling or a staggering potential, all realizable in 
optical lattices \cite{Goldman2010a}. The physics caused by these additional 
effects are indeed very interesting and leaving them out may seem quite 
restrictive, but the results we discuss in this article are quite general and 
require only that the states are topological and do not rely on the detailed 
nature of the edge states. 

\begin{figure*}[t]
\includegraphics[width=\linewidth]{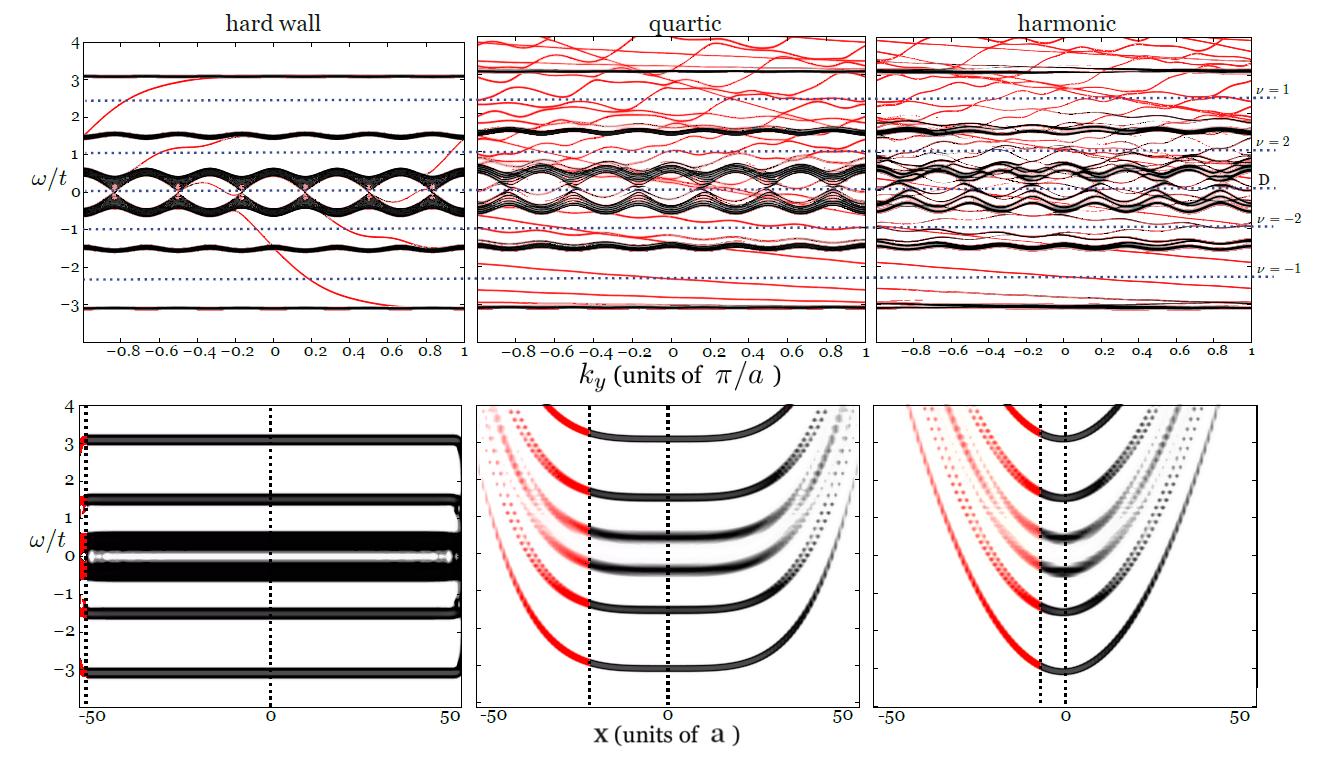}
\caption{(Color online) Integrated spectral function for a system described by 
	\eqref{Hamiltonian1} with flux $\alpha=1/6$ and stripe geometry with 
	$100\times100$ lattice sites \changed{\cite{CylinderGeometry}}. Three experimentally relevant confinements of 
	the form $V(x)=V_0 (x/L)^\delta$ are shown: a) hard wall, $\delta \rightarrow 
	\infty$, b) quartic confinement, $\delta = 4$ and c) harmonic confinement, 
	$\delta = 2$.  The spectra in the upper row, $\rho_L(k_y,\omega)$, show the 
	$k_y$-dependence along the periodic direction, integrated over the left half 
	of the confinement direction, see text, and the real space spectra of the 
	lower row, $\tilde{\rho}(x,\omega)$, show the $x$-dependence along the 
	confinement direction. To the right of the figure are the transport 
	coefficients, calculated using \eqref{transcoeff} with the Fermi edge set to 
the corresponding dotted line. For hard wall and quartic confinement there is an appreciable number of bulk bands and the edge states are clearly distinguishable, whereas within harmonic confinement we consider almost all of the states to be edge states. In each case we indicate left edge states in red (gray), while the 
remaining bulk states are shown in black, corresponding to the left half of the cylinder in real space, see lower plots.
Two possible approaches exist for designation of edge and bulk states in softly confined systems. As seen in the upper row, energy regions with well defined topological quantum numbers can be identified in the spectrum. The corresponding states can be designated as part of the edge, and the remaining ones as the bulk. Alternatively, we can define the edge as the point at which no states have energies within the range of energies covered by states at the center of the trap. We use the latter designation, although there is little difference between the two methods.
}
\label{fig:a16_hqh}
\end{figure*}

\section{Edge states in cylindrical geometries}\label{sec3}
The defining property of topological insulators in a semi-infinite system is the 
emergence of gapless edge states which are localized at one edge and robust 
against perturbations of the system, e.g. potential or magnetic disorder.  
Furthermore, the presence of these states is the origin of the currents measured in QH \changed{\cite{vonKlitzing1980, Delahaye1986, Hatsugai1993}} and QSH samples \cite{Bernevig2006,Kane2005,Koenig2007} which are well known to be strictly quantized when $S_z$ is a good quantum number.  
Topological phases are typically distinguished by the transport properties of 
the edge states, specifically by the quantized charge (or mass for neutral 
atoms) that is transported at a single edge \cite{Kohmoto1989,Hasan2010}. One method of determining the 
topological quantum number for a given system is therefore to calculate the 
energy spectrum on a cylindrical geometry and to evaluate the transport 
properties of the edge states directly.  Alternatively, one may determine the 
topological quantum numbers from the dispersion relation of filled bands \changed{\cite{Gurarie2011} or the corresponding eigenstates \cite{Hatsugai1997,Thouless1982,Hasan2010}} in the 
corresponding infinite system.

In this section we will focus on a cylindrical geometry and determine the spectral 
functions of the system of interest via exact diagonalization of the Hamiltonian 
for a finite system of size $100\times100$ \changed{\cite{CylinderGeometry}}. We discuss the properties of the 
edge states of the system by analyzing the integrated spectral function in 
quasi-momentum space and real space for several kinds of boundaries and show the 
robustness of the edge states against perturbations by switching on a disordered 
potential.

\subsection{Identification of topological invariants}
Topological phases can be characterized either by analyzing the band structure properties of 
the infinite system, or in terms of the transport properties of the system in a 
confined geometry. While the first approach is insensitive to the specific shape 
of the confining potential, the latter may in principle strongly depend on these details. In 
this section we discuss the properties of edge states at an infinite wall 
boundary, realized by open boundary conditions at the edges of the cylinder, 
henceforth referred to as stripe geometry.  For this kind of boundary the 
topological quantum number of the infinite system is equivalent to the 
transport coefficient $\nu$ of the finite system, a relation known as the 
bulk-boundary correspondence \cite{Hatsugai1997, Gurarie2011}. The coefficient $\nu$ counts 
the difference in number of forwards-moving and backwards-moving states at the 
Fermi edge, which represents the net transport for low-energy excitations and 
hence the quantized edge current $I_E$ \cite{Hatsugai1993}.  Explicitly we have \cite{QN}
\begin{equation}
\nu_m=\sum_{\alpha_m}\mbox{sign}(\partial_{k_y} \epsilon_{\alpha_m}(k_y)),
\label{transcoeff}
\end{equation}
where $\alpha_m$ labels \changed{the states at the Fermi edge with energy 
$\epsilon_{\alpha_m}(k_y)=\epsilon_F$} and $m=L,R$ for the left and right edge, 
respectively. Eq. \eqref{transcoeff} can be obtained by applying the well-known Laughlin argument
to a cylindrical geometry and subsequently following the procedure described in \cite{Hatsugai1997}, where 
no details of the trapping potential are required. For the gauge $\mathbf{A}=(0,Bx,0)$, the single particle 
Hamiltonian \eqref{Hfirst} obeys the symmetry 
\changed{$H(\mathbf{x},\mathbf{p})=H(-\mathbf{x},-\mathbf{p})$}, which leads to 
$\nu_L=-\nu_R$. Throughout this article we will only consider the \changed{Hall transport 
coefficient for the left edge $\nu \equiv \nu_L$, which is identical to the 
topological $\mathbb{Z}$ quantum number of the infinite system and determines the Hall conductance $\sigma_{xy}=\nu e^2/h$}.  The topological 
$\mathbb{Z}_2$ quantum number $\nu_2$, which indicates QSH phases in the 
corresponding spin-1/2 system \cite{Goldman2010a} can then be obtained, if $S_z$ 
is a good quantum number, by \cite{Wu2006,Xu2006}
\begin{equation}\label{QSHind}
\nu_2=|\nu|\mod 2.\end{equation}
If $\nu_2 = 1$ the system will exhibit a QSH phase.

In this paper we make an explicit distinction between 
	an ``edge state'' and an ``edge mode''. An edge state always refer to an 
	eigenstate of the Hamiltonian that is localized to one edge, whereas an edge 
mode refers to a series of edge states that are smoothly connected in 
momentum-space.  Although this distinction is not necessary for hard-wall 
systems, it is required for soft boundary systems.

\subsection{Cylindrical geometry with open boundary conditions}
We first consider a system described by \eqref{Hamiltonian1} with a step potential $V$ 
that is zero for $|x|\leq L_x/2$ and infinite elsewhere, with $L_x$ sufficiently 
large, providing a hard-wall boundary at the edges of the cylinder.

Since the quasi-momentum in the $y$-direction is a well-defined quantum number 
and we are interested in transport coefficients for this direction, it is 
convenient to represent the spectrum of the system in terms of the integrated 
spectral density $\rho_L(k_y,\omega)\equiv\int_{-L/2}^0 dx\rho(x,x,k_y,\omega)$, where the spectral
function is defined as
\begin{equation}
\rho(x,x',k_y, \omega)=-2\mbox{Im}\langle x,k_y|\frac{1}{\omega-H+i0^+}|x',k_y\rangle,
\end{equation}

\begin{figure}[t!]
\includegraphics[width=\linewidth]{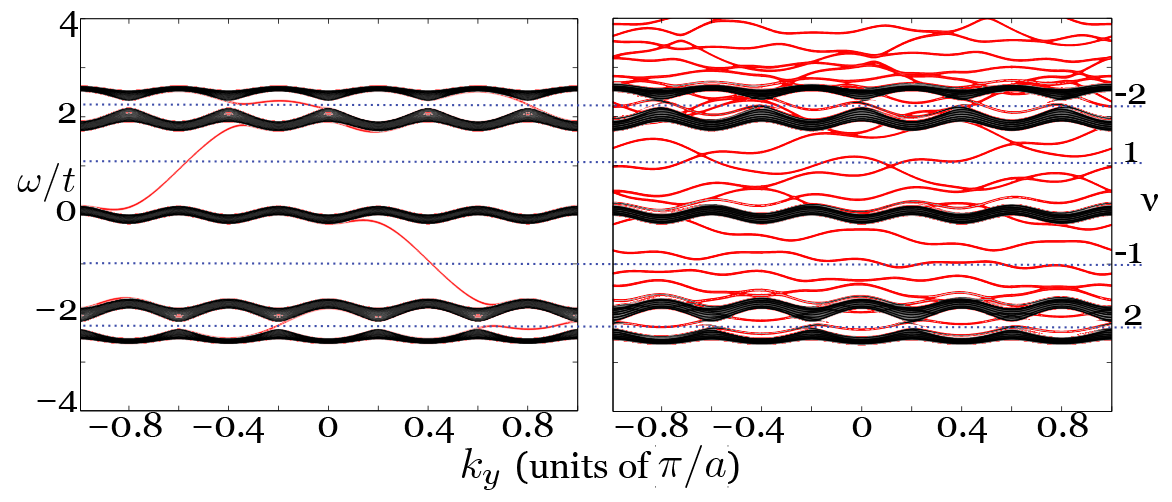}
\caption{(Color online)  Integrated spectral function $\rho_L(k_y,\omega)$ for a 
system described by \eqref{Hamiltonian1} with flux $\alpha=2/5$ and hard-wall 
(left) and quartic (right) confinement. Edge states are shown in red (gray), while bulk states are shown in black.}
\label{fig:a25_inv}
\end{figure}

We integrate only over the left half of the system in real space, so 
as to separate the left from the right edge states \cite{Remark5}. In Fig.~\ref{fig:a16_hqh}a), 
upper, the integrated spectral density $\rho_L(k_y,\omega)$ is shown for 
$\alpha=1/6$.  One can identify the bulk states which are grouped into six thick 
bands and the edges states, which close the gaps between the bands. To determine 
the transport coefficients and possible topological phases, we place the Fermi 
edge in a bulk gap and apply \eqref{transcoeff} to the dispersion of the 
edge states.

\begin{figure*}[t]
\includegraphics[width=\linewidth]{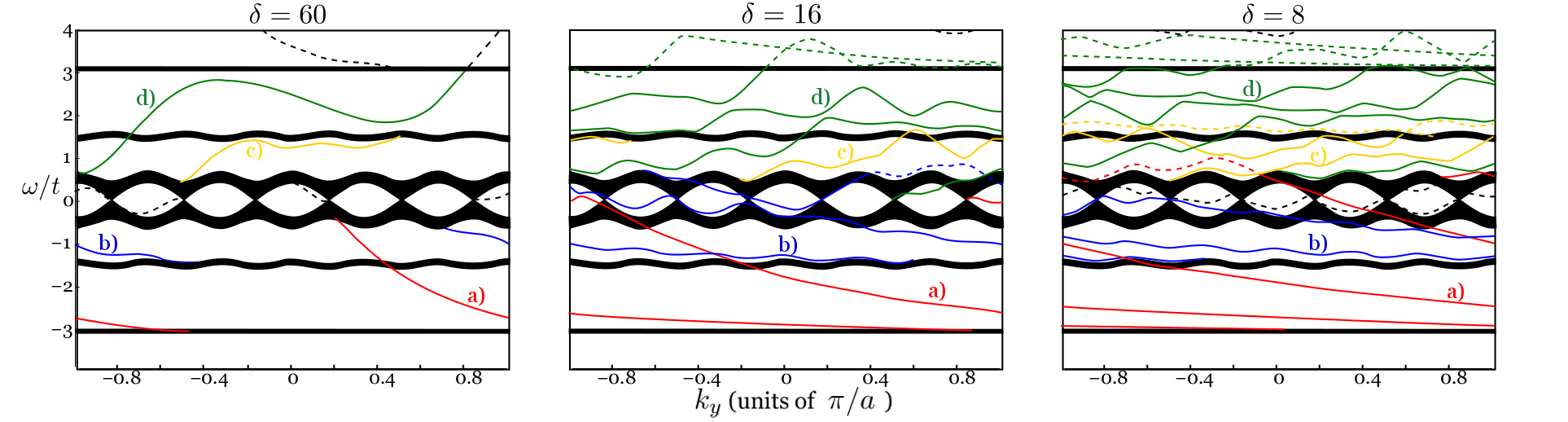}
\caption{(Color online) False color diagram of the integrated spectral density 
	$\rho_L(k_y,\omega)$ for a system described by \eqref{Hamiltonian1} with 
	flux $\alpha = 1/6$, stripe geometry, and a confining potential, $V(x) = V_0 
	(x/L)^\delta$.  Bulk bands are indicated in black and edge modes as colored 
	curves, \changed{also marked as $(a), (b), (c), (d)$}. There exist several true crossings and avoided crossings in the 
	spectra which combine to preserve the topological invariants for any 
	confinement exponent $\delta$. Auxiliary states of the corresponding edge 
	modes are shown with dashed curves. The auxiliary states do not influence 
the topological phases of the system, since they always come in pairs with opposing 
velocities.}
\label{fig:a16_modes}
\end{figure*}
There are several phases visible in this system. If the Fermi edge lies 
within a bulk band, the system is in a trivial metallic phase. If the Fermi edge lies 
between the third and fourth band, there are a set of Dirac points with a linear dispersion 
and the phase is a semi-metal. In the bulk gaps, which lie between the other bands, the 
edge modes place the system in a quantum Hall phase with $\nu=-1$ and $-2$ for the first and 
second bulk gap, respectively, and inverted for the \changed{third, fourth} bulk gap. In shorthand, we can 
specify the phases between the bands by $\mathrm{gap}_{1/6} = \{-1,-2,\mathrm{D},2,1\}$, 
where D represents Dirac points. In the analogous spin-1/2 system, $\nu=\pm1$ indicates a 
QSH phase whereas $\nu=\pm2$ corresponds to a normal insulator, due to lack of topological 
protection.

In addition we also investigate the case where $\alpha=2/5$, see Fig.~  
\ref{fig:a25_inv} for which we find $\mathrm{gap}_{2/5} = \{2,-1,1,-2\}$. Note 
that there are no Dirac points for $\alpha=2/5$. The differences between 
gap$_{1/6}$ and gap$_{2/5}$ appear as different real space behaviors within 
soft confining potentials that are not visible within a hard-wall confinement, 
as we demonstrate in the next section.

\subsection{Cylinder with soft boundaries}

With soft boundaries, it becomes relevant to look at the spectra of the system 
in real space along the $x$ axis, $\tilde{\rho}(x,\omega) = \int dk_y\, 
\rho(x,x,k_y,\omega)$, as well as the partially integrated spectra in 
quasi-momentum space along the $k_y$ axis, $\rho_L(k_y,\omega)$. The 
quasi-momentum spectra allows us to extract transport coefficients and discuss 
the dispersion of the edge modes.  On the other hand, the real space spectrum 
exhibits some unusual features for different trapping conditions.  We consider a 
lattice of size $100\times100$ and trapping geometries of 
the form $V(x) = V_0 (x/L)^\delta$ where $V_0=10t$ and $L=50a$ is chosen such that 
$V(x = 50a) = V(x = -50a) = 10t$ which is larger than the energy spanned by the infinite 
system $\sim 8t$.  Three particular values of $\delta$ are relevant to experiment: 
$\delta \rightarrow \infty$ which reproduces hard-wall boundary conditions, $\delta=4$ for 
quartic confinement and $\delta=2$ for harmonic confinement.  In most optical lattice 
experiments, the confining potential is a result of the Gaussian envelope of the 
finite beam width of the lasers and in the center of the trap we may 
approximate this confinement by its leading order harmonic term.  However, it is 
has been suggested \cite{Gygi2006} that one can remove the harmonic term by 
superimposing an anti-trapping Gaussian beam of different detuning to the 
trapping beam, which then promotes the quartic term to the leading order 
approximation of the trapping potential, i.e.  $V(x) \propto x^4$. This scheme was realized 
in optical lattice experiments to improve quantum phase diffusion experiments 
\cite{Will2010, Will2011}. We investigate these trapping geometries below in further detail.

\subsubsection{General features and preservation of topological invariants}
In Fig.~\ref{fig:a16_hqh}, we show a comparison between $\rho_L(k_y,\omega)$ and 
$\tilde{\rho}(x,\omega)$ for $\alpha=1/6$ and hard wall, quartic and harmonic 
confinements that are relevant to experiment. One can see that the potential does not 
gap the system, and edge modes continue to connect the bands. The transport 
coefficients of the soft boundary systems, indicated to the right of 
Fig.~\ref{fig:a16_hqh}, are insensitive to the trapping potential. In other words, there 
exist energy ranges in which we can identify a transport coefficient, which is identical 
for all confinements we consider.

Comparing  the $k_y$-dependent soft-boundary spectra  of  the upper row in 
Fig.~\ref{fig:a16_hqh},  we make two observations: 1) we can readily identify 
highly degenerate regions of bulk bands in the hard-wall and quartic 
confinements, and 2) we find that the dispersion of edge modes that are present 
within quartic confinement do not change noticeably when changing the 
confinement to the harmonic trap.  In contrast, the rest of the spectrum is 
significantly modified, such that the ratio of bulk to edge states is very small.
To define such a bulk region in the soft boundary system, we assume the edge begins at 
a distance from the trap center where none of the states at this point overlap in energy 
with any of the states in the very center of the trap (i.e. at $x=0$).

Analogously, we can clearly identify a bulk region from the 
$x$-dependent spectra in the bottom row of Fig.~\ref{fig:a16_hqh} for the 
quartic trap, but not in the case of the harmonic trap.    From this we conclude 
that the quartic trap is likely the best trapping potential for observing 
effects of both the bulk system and topological edge states in an experimental 
setup, if it is not feasible to artificially implement hard wall boundaries as 
proposed, for example, in \cite{Goldman2010a}.  Furthermore, we observe no 
overlap between states of different edges, which has been proposed to destroy 
edge states via couplings between the edges \cite{Stanescu2009}. This again 
shows that the edge states are topologically protected and robust against 
external changes in the potential.

In Fig.~\ref{fig:a25_inv}, we also show a comparison of $\rho_L(k_y,\omega)$ for 
the case where $\alpha = 2/5$ between a hard wall (left column) and a quartic 
confined system (right column). We again calculate the transport coefficients of 
both systems and list these next to the plots to show that these also coincide 
for all trapping potentials.

To better understand the details of the rather complicated spectra of the 
quartic trap, we choose to follow the edge modes of the hard-wall confinement by 
smoothly varying the trapping exponent $\delta$.  We show plots for $\delta = 
60, 16, 8$ and $\alpha=1/6$ in Fig.~\ref{fig:a16_modes}, where we have 
artificially colored the spectra to indicate each edge mode. $\delta=60$ 
represents a very steep trap, and is almost identical to the hard-wall case: 
with the color designation, one sees that the blue \changed{(marked as $(b)$)} and yellow \changed{$(c)$} edge states are 
only present in the 2nd and 3rd bulk gaps, respectively, whereas the red \changed{$(a)$} and 
green edge states \changed{($d$)} span two bulk gaps.  As the confinement is made softer, we see 
that an edge mode may cross the BZ more than once, and that the energy range of 
the edge states changes, e.g. with $\delta=8$ the red state \changed{($a$)} now extends into the 
3rd bulk gap.  However, whenever this occurs, the state forms an avoided 
crossing at some higher energy with a different edge state and is forced 
downwards in energy, a process which preserves the value of the topological invariants. We 
represent this in the false color diagram by a dotted line for parts of the edge 
states that are non-topological, i.e. not connecting different bulk bands.  For 
$\delta=16,8$, we can consistently see this 
occurring in the most energetic edge mode (colored green, \changed{($d$)}), which extends above 
the highest bulk band, and forms an avoided crossing with the non-topological 
edge state created by the effect of the trapping on the highest bulk band.

Note that, due to the trapping potential, several edge states that 
belong to the same edge mode may exist for one value of energy.

\subsubsection{Merging and splitting of edge states}
When the number of edge modes changes, as the Fermi edge crosses a bulk band in 
the hard-wall boundary system, either an edge mode must be created, or an edge 
mode must merge into either the bulk band itself or with another edge mode. In 
the soft-boundary system we can see some very non-trivial behavior that shows 
the complexity of these processes.

\begin{figure}[t!]
\includegraphics[width=\linewidth]{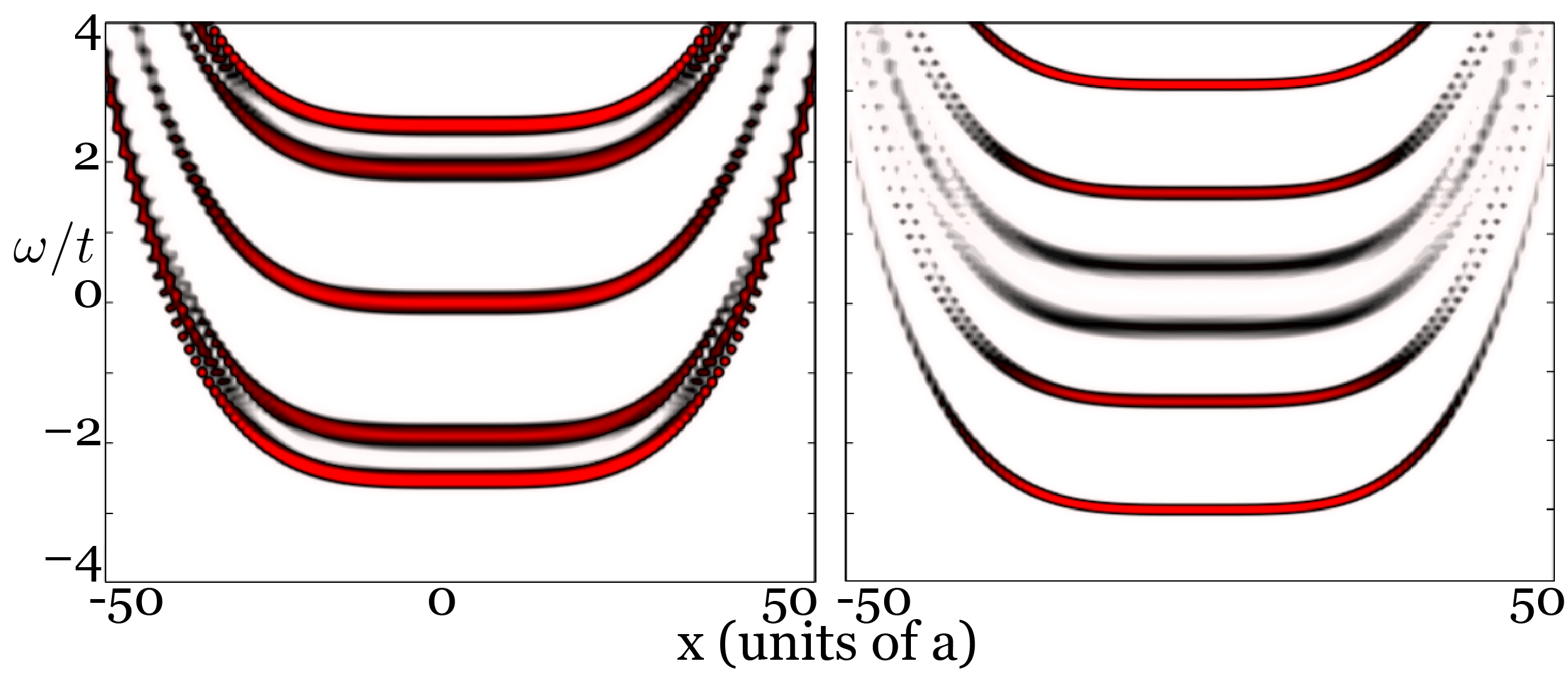}
\caption{(Color online) Integrated spectral density $\tilde{\rho}(x,\omega)$ for a system 
described by Eq. \ref{Hamiltonian1} on a $100\times100$ lattice with quartic confinement 
$V$. In the left figure, the flux is chosen to be $\alpha=2/5$, while in the right figure 
$\alpha=1/6$. In the left figure, the edge states leaving the second bulk band immediately 
merge with the edge states of the first bulk band to form a state spatially localized 
between the bulk bands with the correct transport coefficient. In the right figure, the 
states leaving the second bulk band split up, i.e. they localize to more than one point in 
space. The inner part of these states merges with the edge states of the third band at 
higher energies. Similar behavior is observed for the $3$rd, $4$th and $5$th band. As 
pointed out in the text, this non-trivial behavior is an indication of the topological 
origin of the edge states.}
\label{fig:aboth_real}
\end{figure}

We first focus on the real space spectra of the $\alpha=2/5$ flux system under 
quartic confinement, see Fig.~\ref{fig:aboth_real}. In the lowest gap, we see 
that two different edge modes, which evolve between the first and second bulk band, merge 
into a single edge mode, which evolves between the second and third bulk band.  In the 
hard wall system, this mode is localized to a single site in the $x$-direction and 
can only be observed in quasi-momentum space.  In the quartic trap, the edge 
states leaving the first bulk band follow the shape of the quartic potential and 
one may expect the same for the states leaving the second bulk band. As one sees 
in Fig.~\ref{fig:aboth_real}, this is not the case.  The states leaving the 
second bulk band immediately start to merge with the edge states from the first 
bulk band and the result is only a single mode at each edge, evolving between 
the second and third bulk band\changed{. Although it is not possible to  
determine topological invariants from real-space spectra, we can link this 
merging behavior to the $k_y$-space spectra of Fig.~\ref{fig:a25_inv} and see 
that it leads to} the correct topological quantum number $\nu=-1$. The same 
effect is again observable between the fourth and fifth bulk band.  
Interestingly, the merging of these modes does not take place via a simple 
overlap of the states, but a gap in real space with negligible spectral weight 
exists between the states originating from the bands and the newly-formed edge 
mode.

In the $\alpha=1/6$ flux system, we also see the opposite effect: the splitting 
of a single bulk band, to connect edge modes of different bands which are 
energetically well separated. In Fig.~\ref{fig:a16_hqh}, the integrated spectral 
density $\tilde{\rho}(x,\omega)$ shows that the \changed{modes leaving the 
second and fifth bulk bands each split into two curves, where a single 
eigenstate has large amplitudes on two spatially separated lattice sites. We 
interpret this splitting as a process that facilitates the connection between 
different bands which we observe in Fig.~\ref{fig:a16_modes}. For example, the 
outer part of the mode leaving the second bulk band can be seen to merge at 
higher energies with the mode that is a product of the third and fourth bands.  
This connection between the bands is analogous to the avoided crossings that we 
observe in the $k_y$-dependent spectra in Fig.~\ref{fig:a16_modes}.}  This very 
non-trivial behavior of modes within the outer region of the system, 
\changed{ combined with transport coefficients which are identical to the topological 
quantum numbers, given by the transport coefficients of the infinite system,} indicates 
that the soft edge states are of topological origin.  To further verify this, we address 
in section \ref{sec:robustness} the robustness of these states against perturbations in 
terms of a disordered background potential.  

\subsubsection{Relation of edge states and bulk bands}
When we look more closely at the dispersion of the edge modes, we can see an 
interesting connection to the bulk bands of the system. We focus on the 
quasi-momentum spectra for the case $\alpha = 1/6$ shown in Fig.~\ref{fig:a16_modes} for 
increasing confinement exponent $\delta$.

The dispersion of an edge mode leaving a given bulk band can be described on two 
different quasi-momentum scales. For a small range of $k_y$, the dispersion 
mimics that of its associated bulk band, and this behavior becomes more 
prominent for smaller $\delta$. This can be seen for the lowest edge modes, colored 
red \changed{($a$)} and blue \changed{($b$)}, e.g. the red \changed{($a$)} mode has a locally flat dispersion, mirroring the flatness of 
the lowest band.  However, when avoided crossings have occurred, such as for the 
yellow \changed{($c$)} and green \changed{($d$)} edge modes, the dispersion of an edge mode cannot simply be 
described by one band alone and corresponds to a mixture of bands.

On the other hand, considering the Brillouin zone as a whole, the edge modes 
become more flat in momentum space, the smoother the confining potential is in 
real space.  This flattening is a direct result of the number of accessible sites 
at the edge. The number of lattice sites, $n_\mathrm{edge}$, that are available 
for an edge state between e.g. the first and the second bulk band, is the number 
of sites $i$ that fulfill \changed{$\epsilon_1-\epsilon_0 \lessapprox V(x_i ) \lessapprox 
\epsilon_2-\epsilon_0$, where $\epsilon_1$ ($\epsilon_2$) are the maximum (minimum) 
energies of the first and second bulk band, respectively, and $\epsilon_0$ is the minimum 
energy of the first bulk band \cite{Stanescu2010}}. In the hard wall system, 
$n_\mathrm{edge} = 1$ but becomes larger the smoother the confining potential 
becomes in real space.  An interesting result for the $\alpha = 1/6$ flux 
per plaquette, is that the flatter the potential becomes, the flatter the lowest 
gap edge modes become, with a corresponding increase of the effective mass of system's 
excitations:
\begin{equation}
m^*\equiv \left(\frac{\partial^2}{\partial k_y^2}\epsilon(k_y)\right)^{-1}\rightarrow \infty.
\end{equation}
This is generally true for soft confinements, as pointed out in 
\cite{Stanescu2010}, but the edge state structure in the case $\alpha=1/6$ 
allows this feature even for relatively steep potentials.

\subsection{Robustness of soft edge states in stripe geometries}
\label{sec:robustness}
One of the most important properties of topological edge states is their  
robustness against even large perturbations, which leads to clearly detectable 
quantized Hall conductance in impure experimental setups. In optical lattice 
experiments	which are, by construction, very clean realizations of condensed 
matter Hamiltonians, perturbations such as disorder are usually not an issue. 
However, since disordered potentials can be implemented in a controlled manner 
\cite{Billy2008,Fallani2007,SanchezPalencia2010}, 
it is of interest to thoroughly investigate how robust the edge 
states are against these kinds of perturbation.  Here, we address this question 
for soft boundaries. The general argument, which illustrates the robustness of 
edge states in condensed matter systems, is the lack of possible backscattering 
processes \cite{Hasan2010}. Counter-propagating edge states are localized on opposite edges 
of the system and are very well separated spatially. Therefore, in huge 
condensed matter systems, these states have vanishing spatial overlap and 
backscattering from impurities is completely suppressed.
In contrast, in finite systems different edge states from opposite edges will 
have a finite overlap in real space, which theoretically allows for 
backscattering processes, and therefore disorder may lead to the opening of a gap 
in the spectrum. However, as we will see from our numerical results, even in 
very small systems ($\approx60$ lattice sites in $x$-direction) this effect is 
not observable.

To verify the robustness of the soft edge states numerically, we perturb the 
system described by Eq. \ref{Hamiltonian1} by adding a disordered background 
potential \begin{equation}\label{Pert}
V_{\mbox{\tiny disorder}}=\sum_l \Delta_l c^{\dagger}_lc^{\phantom{\dagger}}_l,
\end{equation}
where $\Delta_l$ is distributed randomly, either by a binary distribution, 
$\Delta_l\in\{0,\Delta_{\mbox{\tiny max}}\}, \forall l$, or by a uniform 
distribution, $\Delta_l\in [0,\Delta_{\mbox{\tiny max}}], \forall l$ \changed{\cite{Disordered}}. For all 
realizations, we found that the edge states stay robust and still connect the 
different bulk band regions without opening a gap up to disorder strengths of 
about $\Delta_{\mbox{\tiny max}}\approx0.5t$ for binary disorder and even larger 
strengths for uniform disorder.

\begin{figure}[t!]
\includegraphics[width=\linewidth]{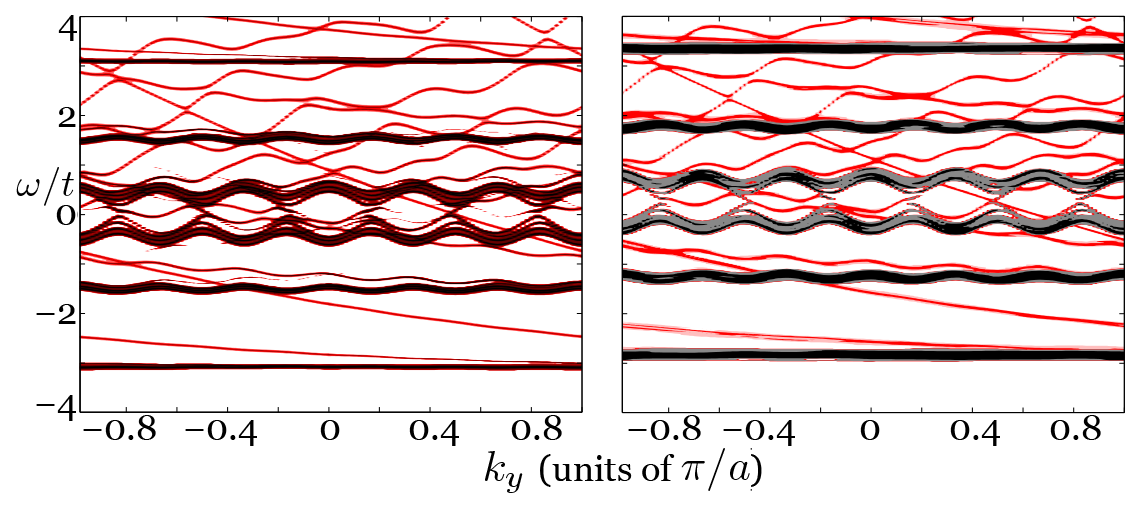}
\caption{(Color online) Integrated partial spectral density $\rho_L(k_x,\omega)$ 
of a system described by Eq. \ref{Hamiltonian1} on a $120\times60$ lattice \changed{with 
$\alpha=1/6$. Right: with an 
additional binary disordered potential, given by Eq. \ref{Pert} and 
$\Delta_{\mbox{\tiny max}}=0.5t$. Left: with the disordered potential being set to zero}. 
The robust edge states are still clearly 
pronounced and gapless, while the former bulk bands are smeared out and show a 
mobility gap (not shown here but obtainable from the Anderson-localized bulk 
eigenstates).}
\label{fig:disorder}
\end{figure}

For example, in Fig.~\ref{fig:disorder} the integrated partial spectral density 
$\rho_L(k_y,\omega)$ is shown for a $120\times60$ lattice system with uniformly 
distributed $\Delta_l$ and $\Delta_{\mbox{\tiny max}}=0.5t$. There is clearly no 
gap in the spectrum and although quasi-momentum is no longer a good quantum 
number, the edge states in momentum space are very sharply centered around a particular value of 
$k_y$ and remain delocalized in the $y$-direction as they were for the 
system without disorder \cite{Remark3}. In contrast, some of the bulk states now 
consist of many quasi-momentum components (not shown in our figure) and therefore become localized to a 
region much smaller than the system size, which can be termed Anderson 
localization.

We have also addressed larger systems with larger boundary regions. These 
systems contain more and more edge states in a given bulk gap, which may 
possibly lead to different backscattering processes between edge states 
located at the same edge and therefore open gaps in the spectrum after 
disorder is introduced. To exclude these possibilities, we studied system sizes 
of up to $60\times240$ lattice sites without finding any indication of gaps in 
the spectrum or localization of the edge states up to disorder strengths of $\Delta_{\mbox{\tiny max}}=0.5t$.

\section{Detection methods}\label{sec4}
So far we focused on a semi-infinite system with stripe geometry. However, 
realistic systems in optical lattice experiments are confined to a finite region 
in all dimensions by the finite beam width of the lasers. In this section, we 
determine the spatial wave functions of a 2D system trapped in both the $x$- and 
$y$-directions and discuss possible detection methods of the resulting edge 
states.
\subsection{Eigenstates of the completely trapped system}
We determine the eigenstates of a system with a confining potential $V$ that 
varies in the $x$- and $y$-directions
\begin{equation}\label{totrap}
V(x,y)=V_0 \left[\left(\frac{x}{L}\right)^{\delta}+\left(\frac{y}{L}\right)^{\delta}\right].
\end{equation}
The parameter $\delta$ determines the shape of the trap and the possible 
eigenstates.  For $\delta\to \infty$ the system is again confined by hard walls 
in both directions, while for $\delta=4$ and $\delta=2$ the system is in a 
quartic and harmonic confinement, respectively. For harmonic confinement, we 
expect the eigenstates that are extended over various lattice sites to be 
circularly symmetric, whereas in the quartic case the potential is no longer circularly symmetric and the states take on a shape that is sometimes referred to as a 
squircle \cite{Makogon2012}.  For this analysis we again will restrict ourselves to a system with 
$100\times100$ lattice sites and a trapping potential with a minimum value of $V_0=10t$ 
along each edge of the lattice.  We again focus on $\alpha=1/6$.

\begin{figure}[t!]
\includegraphics[width=\linewidth]{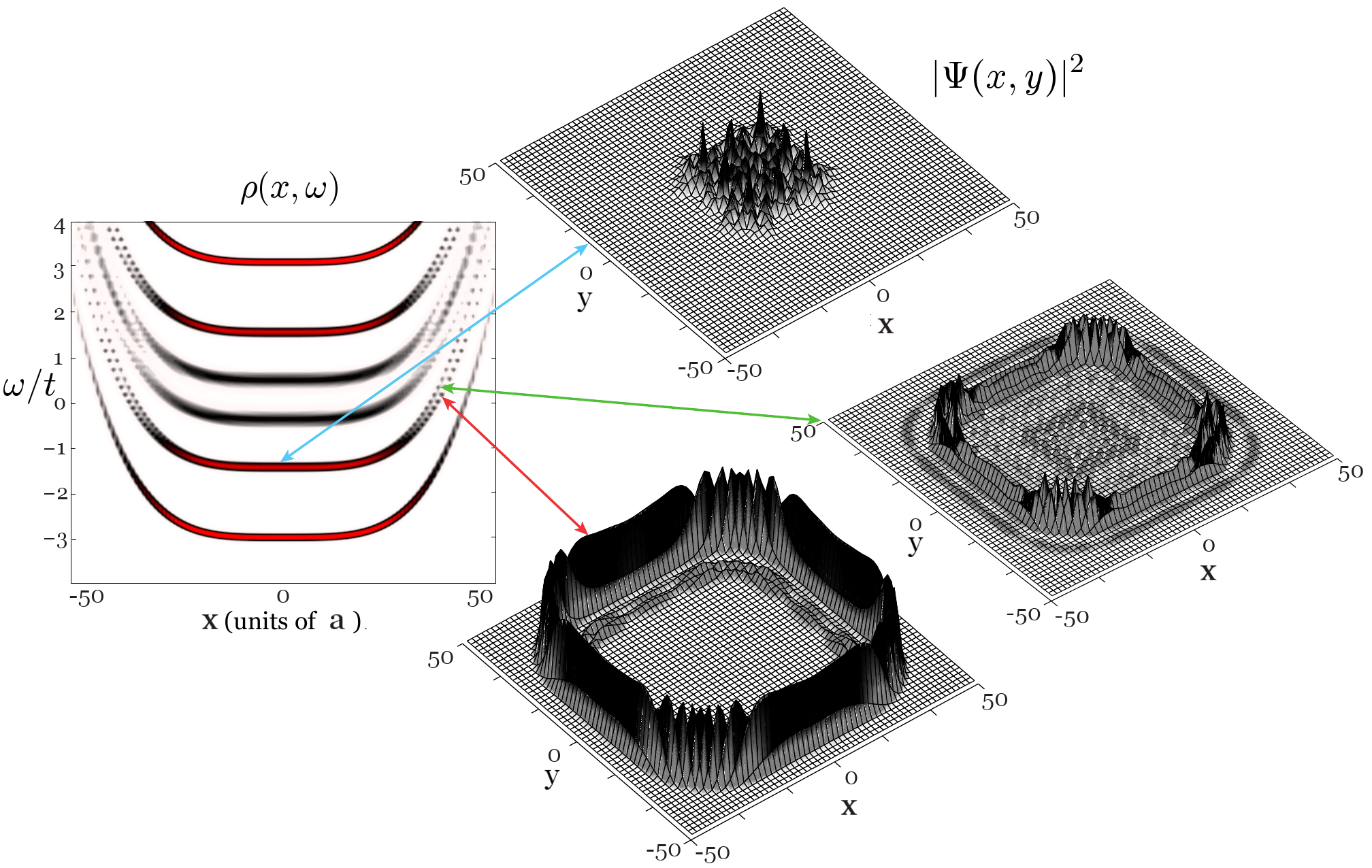}
\caption{(Color online) Wave function $|\Psi(x,y)|^2$ \changed{as a function of the lattice spacing $a$} 
of different eigenstates 
of the system within complete quartic confinement. The real space spectral 
density shown to the left is a cross section of the complete system: 
$\rho(x,y=0,\omega)$. Three particular eigenstates have been shown, and their 
energies indicated by arrows to the spectral density. The wave functions belong 
to a) a bulk region and b) and c) to a pair of edge states splitting up after 
leaving the second bulk band.}
\label{fig:a16_compconf}
\end{figure}

In Fig.~\ref{fig:a16_compconf}, we clearly see the different real space 
distribution of edge states compared to bulk states.  The bulk states are 
delocalized over a region of about $30\times30$ lattice sites, while the edge 
states for a given energy follow the isolines of the quartic potential, and are 
strongly confined to these regions.  Comparing figures b) and c) for the 
completely trapped system one can see the splitting of the edge states leaving 
the second bulk band as the two states have a weak overlap with one another. The 
shape of the edge states in the quartic confinement looks similar to that which 
one would expect in the hard wall system (like those explicitly shown in 
\cite{GoldmanarXiv}) and differs only at the very corners of the system.  We therefore expect 
similar single particle excitations for the quartic confinement as for the hard 
wall confinement when probing the edge states in experiment.

\begin{figure}[t!]
\includegraphics[width=\linewidth]{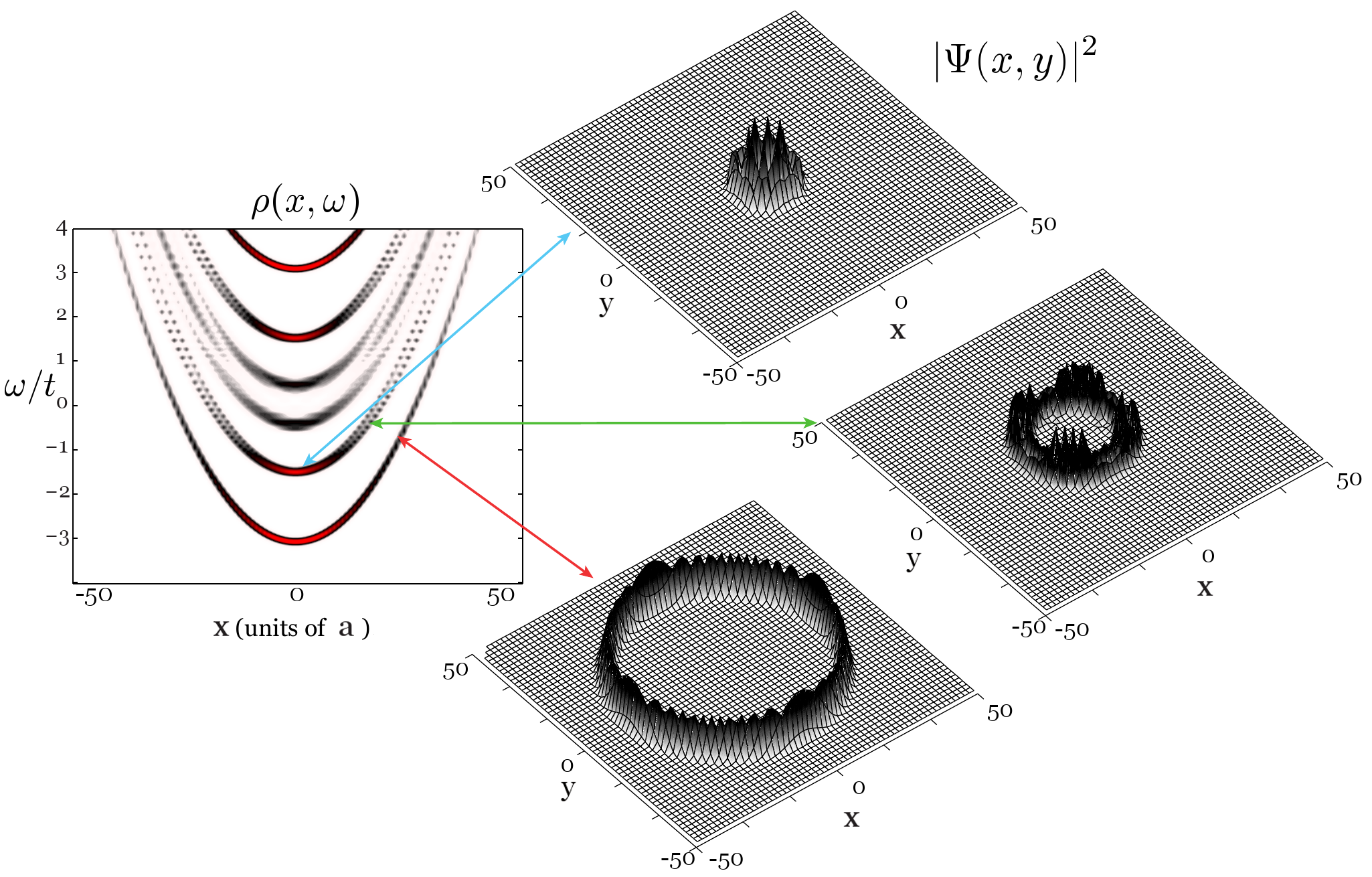}
\caption{(Color online) As in figure \ref{fig:a16_compconf} but with complete 
harmonic confinement. The three states are shown from a) a bulk region and b) and c) 
edge states of states belonging to different bulk bands. \changed{Spatial coordinates in units of the 
lattice spacing $a$.}}
\label{fig:a25_compconf}
\end{figure}

The situation slightly changes when looking at the harmonically confined system.  
There, the confining potential is circularly symmetric and one may expect that 
the eigenstates reflect this symmetry.  The wave functions of the harmonically 
confined system are shown in Fig.~\ref{fig:a25_compconf}. As already seen from 
the spectral density plotted in Fig.~\ref{fig:a16_hqh}, the former bulk region 
is tightly confined to very few lattice sites in the center of the trap, which makes it difficult to define a bulk region in the harmonic trap.  On 
the other hand, the edge states chosen reflect the radial symmetry of the 
trapping potential and again are localized along the isolines of the trapping 
potential.  This already indicates that for the harmonic confinement we expect 
very different excitation dynamics than for the quartic and hard wall confinement, 
where significant parts of the eigenstates are quasi-one-dimensional.

For harmonic confinement, one can solve the continuum model analytically  in the 
absence of the lattice \cite{Messiah1961, Cohen1977} and the resulting wave functions are quite similar to 
those from the lattice calculation. The major difference in the continuum case is that 
no edge states from different bulk bands merge, since the hall conductivity 
always increases by one when passing a bulk band and the different bulk bands 
are not connected. Additionally, angular momentum is only a good quantum number in the continuum case.
\subsection{Bragg spectroscopy}
An important question concerning topological non-trivial phases in ultracold 
atoms is whether the edge states are detectable with existing 
experimental tools. Due to the lack of stationary transport in optical lattice 
experiments, it is not feasible to directly measure the Hall conductance, and 
one has to consider alternative approaches \changed{\cite{Stanescu2010, Scarola2007}}.  Several possibilities for detecting 
edge states or topological quantum numbers in optical lattice experiments have 
been proposed.  Some require careful experimental implementation such as Bragg \cite{Liu2010,Gerbier2012} 
or Raman spectroscopy \cite{Dao2007} and others take advantage of easily accessible observables 
like time-of-flight (ToF) patterns or density profiles.

Density profile measurements were proposed by Umucalilar~\etal~\cite{Umucalilar2008} to directly separate the bulk and edge densities between 
different bands.  However, as already pointed out in \cite{Stanescu2010}, these 
profiles do not show the required structure, as can be seen in Fig.~\ref{fig:a16_hqh} (lower): the bulk bands all occupy approximately the same 
real space extent.  Hence, this method is not applicable to topological systems 
in general.

Alternatively, ToF measurements have been proposed by Zhao \etal~ 
\cite{Satija2011} to exhibit minima and maxima that depend on the topological 
number of the system.  While this is true for the specific cases they were 
investigating and also for our system in the case of $\alpha=1/6$, we found that 
it is not valid in the case of $\alpha = 2/5$ and therefore cannot be reliably 
used as a detection method in experiment. In contrast, Alba~\etal~\cite{Alba2011} propose using ToF measurements as a method to identify skyrmions, by focusing on topological properties of pseudo-spin vectors within the Hamiltonian on the Bloch sphere. However, this method focuses on bulk properties rather than the edge modes that we consider here.

We choose to focus instead on Bragg spectroscopy, which probes the dynamical 
structure factor $S(q,\omega)$ of the underlying system. Bragg scattering of 
topological insulators in optical lattices has been previously considered for 
the case of the quantum anomalous Hall effect \cite{Liu2010}.  However, no 
inhomogeneity of the lattice was considered. Recently, Goldman \etal~\cite{Gerbier2012} have investigated Bragg spectroscopy theoretically, 
considering shaped lasers to probe angular momentum states within circularly 
symmetric traps. While this is a novel implementation to enhance the detection 
of edge states, we demonstrate that one is able to observe edge states using a 
simple linear Bragg coupling which, due to technical limitations, may be the 
only option available to a particular experiment. Furthermore one can observe 
differences in Bragg spectroscopy between the various bands that we show is not 
due to chirality considerations. We do not propose an explicit experimental 
setup and simply assume that one can measure the dynamical structure factor 
directly. One such proposal to measure this precisely in an optical lattice is 
the so called "shelving method" \cite{Gerbier2012}.

When performing Bragg spectroscopy, the system is illuminated by two laser 
beams, described by wave vectors $\bm{p}_1, \bm{p}_2$ and frequencies $\omega_1=p_1 c$, 
$\omega_2=p_2 c$, respectively, and the differences in these quantities, 
$\bm{q}=\bm{p}_1-\bm{p}_2$ and $\omega=\omega_1-\omega_2$, allows for transitions between 
different eigenstates of the original system.  The Hamiltonian describing the 
interaction of the system with the laser beams is then given by 
\begin{equation}\label{BraggHam}
H_{\mbox{\tiny Bragg}}=\frac{\Gamma}{2}\int d^2p\left(e^{-i\omega t}\Psi^{\dagger}(\bm{q}+\bm{p})\Psi(\bm{p})+\mbox{h.c.}\right),
\end{equation}
where $\Psi^{\dagger}(\bm{p})$ is a field operator, creating a particle with real 
momentum $\bm{p}$ and $\Gamma$ is the coupling strength of the lasers \cite{StamperKurn1999,Stenger1999,Steinhauer2002}.

\begin{figure}[t!]
\includegraphics[width=\linewidth]{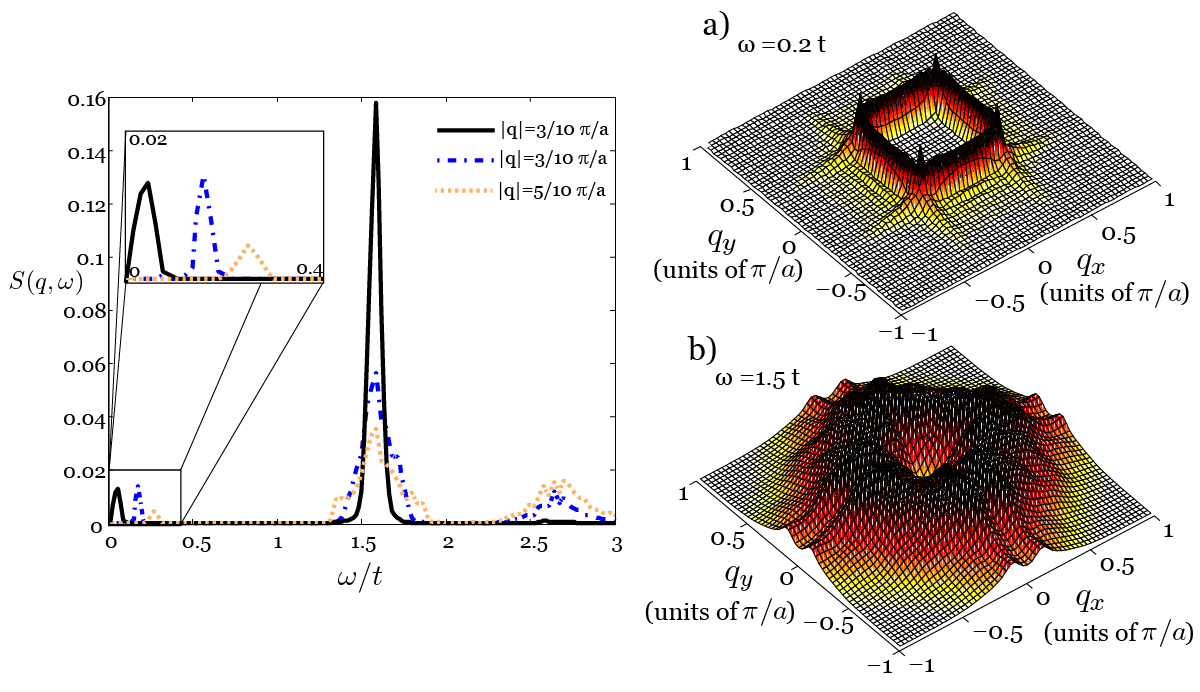}
\caption{(Color online) Dynamical structure factor for a system with quartic 
	confinement and Fermi energy in the first bulk gap ($\epsilon_F=-2t$). Left: 
	$S(q,\omega)$ for fixed momentum $q$ as a function of $\omega$. The first 
	peak belongs to edge$\rightarrow$edge scattering and its position is 
	sensitive to $q$ and can be written, for small $\omega$, as 
	$\omega_q=v_{F,\mathrm{edge}} q$.  The second and third peak belong to 
	edge$\rightarrow$bulk scattering from edge states into the third and fourth 
	bulk bands, located around $\epsilon=0$ and to bulk$\rightarrow$bulk 
	scattering from the first to second bulk bands, where the frequency is 
	independent of $q$. No signal appears of scattering from edge states to the 
	second bulk band, located at $\omega=0.5t$, indicating a disconnection 
between these states, i.e. these states have vanishing matrix elements of the Bragg operator. Right: $S(q,\omega)$ for fixed frequency as a function of 
momentum transfer $q$ for a) edge$\rightarrow$edge scattering at $\omega=0.2t$, 
b) bulk$\rightarrow$bulk scattering processes at $\omega=1.5t$.}
\label{fig:bragg_quartic}
\end{figure}

The dynamical structure factor in linear response theory for an infinite 
homogeneous system is directly connected to the density-density correlation 
function $\chi_{q,q}(\omega)$
\begin{equation}\label{flucdiss}
S(q,\omega)=-\frac{1}{\pi}\mbox{Im} \chi_{q,q}(\omega)
\end{equation}
via the fluctuation dissipation theorem \cite{NegeleOrland}. For our case, we 
have to evaluate $S(q,\omega)$ for the inhomogeneous system, where the 
quasi-momentum is no longer a good quantum number. Within the linear response 
approximation, and accounting for the finite size of the system and finite time 
of the measurement process, we find:
\begin{equation}\label{dynstruc}
S(q,\omega)=|\Gamma|^2\Delta \sum_{\mu,\lambda}  \frac{n_{\lambda}(1-n_{\mu})|A_{\lambda,\mu}(q)|^2}{(\omega-\omega_{\mu}+\omega_{\lambda})^2+\Delta^2}.
\end{equation}
Here, $\lambda, \mu$ label the single-particle eigenstates of the system, 
$n_{\nu}$ and $\omega_{\nu}$ are the occupation number and energy, respectively, 
of the state $\nu$. We introduce a Lorentzian broadening factor $\Delta$, to 
allow evaluation in a system of finite size.  The scattering amplitude 
$A_{\lambda,\mu}(q)$ is the probability of a particle in state $\mu$ to scatter 
into the state $\lambda$ by gaining momentum $q$ and is given by the integral
\begin{equation}\label{scatamp}
A_{\lambda,\mu}(q)=\int d^3r e^{-iqr}\psi_{\mu}^*(r)\psi_{\lambda}(r).
\end{equation}
After determining the single particle eigenstates of the system, we can directly 
calculate the dynamical structure factor. Because we are focusing on the 
detection of edge states, we investigate a system with a Fermi energy located in 
a bulk gap at $\epsilon_F=-2t$ (see Fig.~\ref{fig:a16_hqh}), where an edge state 
is located.  There are now four general scattering processes possible, 
edge$\rightarrow$edge, edge$\rightarrow$bulk, bulk$\rightarrow$edge and 
bulk$\rightarrow$bulk. Edge$\rightarrow$edge scattering is clearly 
distinguishable by analyzing the dynamical structure factor. Given a frequency 
$\omega$, the set of possible momentum transfers allowed to another edge state 
is very limited because the edge states are well localized in momentum space.  
For the case of edge$\rightarrow$bulk scattering, many different momenta are 
accessible and therefore we see a signal regardless of the value of $q$.  This means 
for a fixed momentum transfer $\bm{q}$, $S(\bm{q},\omega)$ as a function of $\omega$ 
consists of a $\delta$-peak approximately around $\omega=qv_F$ \cite{Remark4} 
and a smeared out region, where the bulk bands are located.  This can be seen in 
Fig.~\ref{fig:bragg_quartic} (left), where the first peak indicates 
edge$\rightarrow$edge scattering and the second and third peaks correspond to 
edge$\rightarrow$bulk and bulk$\rightarrow$bulk scattering.

On the other hand, for a fixed frequency $\omega$, the response in momentum space describing 
edge$\rightarrow$edge scattering looks quite different from that obtained from 
edge$\rightarrow$bulk scattering, as one can see from 
Fig.~\ref{fig:bragg_quartic} a) and b), respectively.  For the quartic 
confinement, the edge states form squircles in real space 
(Fig.~\ref{fig:a16_compconf}), which means that low energy excitations are most 
favorable in $x$- or $y$-direction, resulting in the square-like structure of 
$S(q,\omega)$ in Fig.~\ref{fig:bragg_quartic}, which is approximately described 
by $\{q_{\mbox{\tiny{allowed}}}\}=\{(q_x,q_y)\ |\ 
\max\{|q_x|,|q_y|\}\approx q_0=v_F/\omega\}$. In contrast, the dynamical structure 
factor of bulk$\rightarrow$bulk scattering from the first to the second band is smeared out
and depends on the Fermi surface of the occupied level at $\epsilon_F$ and the 
Fermi surface of the unoccupied band $\tilde{\epsilon}_F = \epsilon_F + \omega$.  
The allowed momenta are approximately described by 
$\{q_{\mbox{\tiny{allowed}}}\}=\{(q_x,q_y) \ |\ 
|q_x|+|q_y|\approx \tilde{\epsilon}_F/\tilde{v}_F\}$ and form a rough square 
which is rotated by $\varphi=\pi/4$ compared to the edge$\rightarrow$edge scattering.  
Note that as a result of the structure of $A_{\mu,\lambda}(q)$, where a 
minimal spatial overlap of the two spatial wave functions is needed for 
obtaining a finite scattering amplitude, high frequency 
edge$\rightarrow$edge scattering is exponentially suppressed because the 
presence of the trap causes energetically separated edge states to be localized 
to different distances from the center of the trap. \changed{This does not occur 
in the equivalent hard-wall system.}

\begin{figure}[t!]
\includegraphics[width=\linewidth]{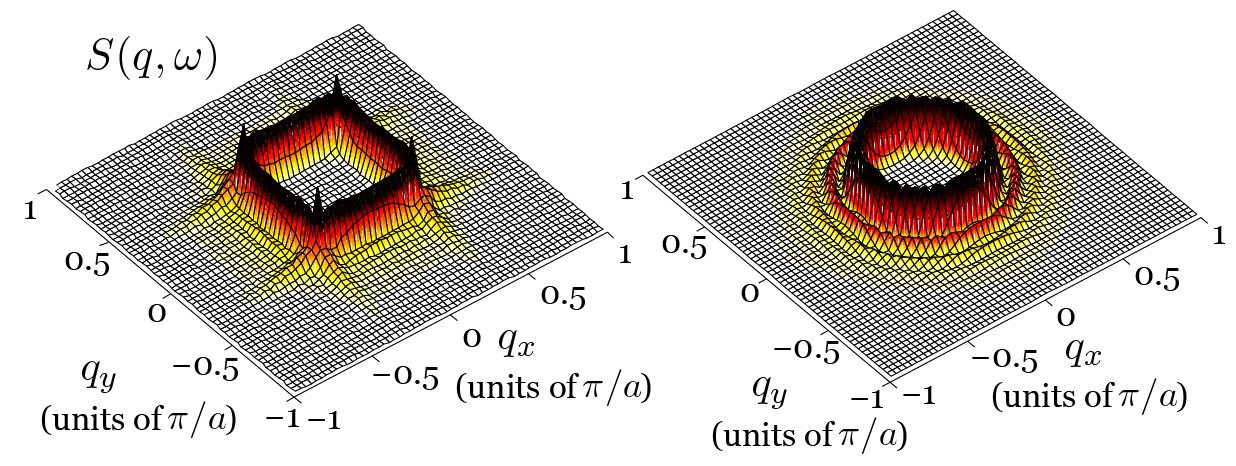}
\caption{(Color online) $S(q,\omega)$ for a fixed frequency $\omega=0.2t$ as a 
function of momentum transfer $q$, for a Fermi energy $\epsilon_F=-2t$. Left: 
$S(q,\omega)$ for the quartic confined system. The system shows a strong 
response when one component of $q=(q_x,q_y)$ has an absolute value 
$|q_{x,y}|=q_0=\omega/v_F$ because excitations along the $x$-axis, $y$-axis are 
most favorable (see Fig.~\ref{fig:a16_compconf}).  Right: $S(q,\omega)$ for the 
harmonically confined system. Here, the response is close to circularly symmetric in 
$q$-space, reflecting the shape of the eigenstates. No particular direction is 
anymore favorable, as long as the absolute value of $|q|=q_0$ is 
fixed.}
\label{fig:bragg_harmonic}
\end{figure}

For the harmonically confined system, $S(q,\omega)$ as a function of $\omega$ for 
fixed $q$ is qualitatively the same as in the quartic system. In contrast, 
$S(q,\omega)$ as a function of $q$ for fixed $\omega$ for edge$\rightarrow$edge 
scattering looks quite different than for quartic confinement. As seen in 
Fig.~\ref{fig:a25_compconf}, the edge states have a circular symmetry and 
therefore no momentum transfer direction is preferred, which leads to the 
set of allowed states forming a circle 
$\{q_{\mbox{\tiny{allowed}}}\}=\{(q_x,q_y) \ | \ \sqrt{q_x^2+q_y^2}\approx \omega/v_F\}$, 
shown in Fig.~\ref{fig:bragg_harmonic}.

\begin{figure}[t!]
\includegraphics[width=\linewidth]{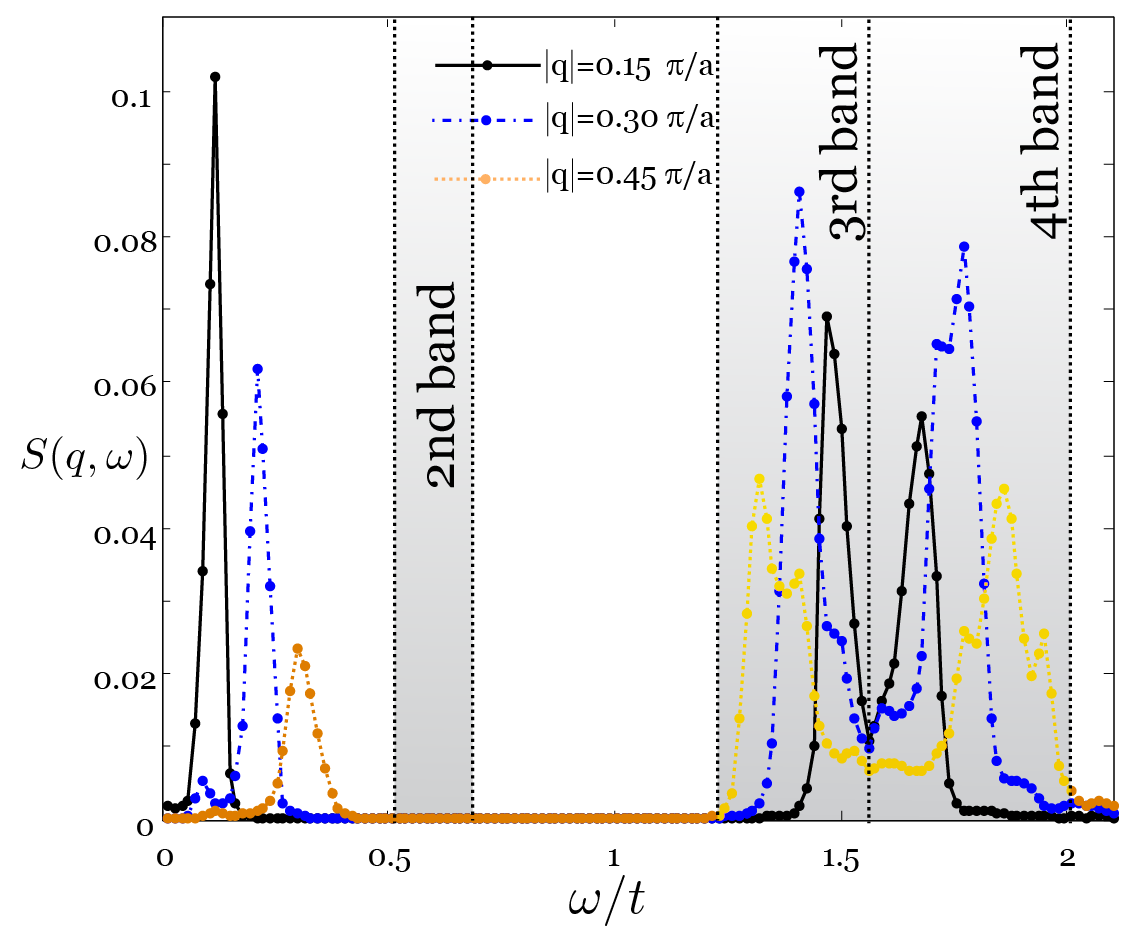}
\caption{(Color online) Dynamical structure factor $S(q,\omega)$ for fixed 
momentum $q$ as a function of frequency, for a Fermi energy $\epsilon_F=-2t$, as 
seen in Fig.~\ref{fig:bragg_quartic} but with artificially suppressed bulk$\rightarrow$bulk 
scattering processes.  The first peak belongs to edge to edge scattering 
processes and is sensitive to the momentum transfer $q$ with approximate 
frequency $\omega_q=v_{F,\mathrm{edge}} q$. The broadened peaks around 
$\omega=1.5t$ belong to edge$\rightarrow$bulk scattering to the third and fourth 
bulk bands.  It is clearly visible that there is no scattering from the edge 
states to the second bulk band, which is located at 
$\omega=0.5t$.}
\label{fig:bragg_edgetoedge}
\end{figure}

An important discovery of our calculations is that there is an obvious absence of 
spectral weight at frequencies where we expect signals of edge$\rightarrow$bulk 
scattering. To highlight this, we calculate an artificial Bragg response 
where we allow only initial states in the energy range $-2.5t < \epsilon < 
-2t$ for transitions to states of higher energy. This means any signal due to 
possible bulk$\rightarrow$edge or bulk$\rightarrow$bulk transitions is 
suppressed. The spectra shown in Fig.~\ref{fig:bragg_edgetoedge} demonstrate 
edge$\rightarrow$edge signal for the first bulk gap, edge$\rightarrow$bulk 
signal to the third and fourth bulk bands, but conspicuously absent signal for 
the edge$\rightarrow$2nd bulk band transitions, which would be expected for 
Bragg frequencies $0.5 < \omega < 1.5$.
This implies that the first edge and second edge/bulk are disconnected, i.e. have a vanishing matrix element of the Bragg operator. 
It is possible to predict this behavior from the dispersion of the 
edge states (see Fig.~\ref{fig:a16_modes}), as one can see that the lowest edge 
mode, colored in red \changed{(a)}, passes unimpeded through the second bulk band, and never 
displays an avoided crossing with the blue \changed{($b$)} edge mode of the second band or the second band itself, while it always merges with the 3rd or 4th band (with which we find non-vanishing matrix elements of the Bragg operator). For higher 
energies, and strong confinements, we see the opposite behavior of avoided 
crossings between red \changed{($a$)} edge modes and yellow \changed{($c$)} edge modes, indicating that one can expect a finite Bragg response from transitions between these states.

\changed{Note that the lack of edge$\rightarrow$bulk scattering is not a result 
of the soft-boundaries inhibiting real-space overlap. We have performed 
equivalent hard-wall boundary calculations where real-space overlap is 
guaranteed but we again observe an absence of signal for disconnected 
edge$\rightarrow$bulk transitions. Note also, that we do not expect to observe a 
clear signal for large frequency edge$\rightarrow$edge transitions regardless of 
the type of trap, as there is larger range of states beneath the Fermi-edge that 
can be accessed with the Bragg laser.  Hence, many different values of $k_y$ 
will contribute, leading to a blurred signal.}

\section{Conclusion}\label{sec5}
In this article, we analyzed the properties of 2D topological edge states in 
softly confined systems with a confinement in one direction of the form $V(x) = 
V_0 (x/L)^\delta$.  By varying the confining potential from a hard-wall to a quartic 
or harmonic potential, we showed that the topological properties of the edge 
states in specific bulk gaps do not depend on the steepness of the confining 
potential, while a confinement sharper than harmonic is required to achieve an appreciable bulk region of the lattice. We suggest that quartic 
confinement is suitable to observe both edge state and bulk properties, which 
may be realized by superimposing attractive and repulsive Gaussian beams. 
Furthermore, we observed the emergence of robust auxiliary edge states, which provide additional structure to edge modes but do not influence the topological quantum numbers.
The main feature of these auxiliary states is that they connect edge states which are spatially separated to bulk bands of the system. This provides a mechanism to preserve topological 
invariance, as soon as the edge states and bulk bands become spatially separated. In these cases the band-structure exhibits a series of avoided crossings that act to preserve the topological invariant. An analysis of the spectral density of softly confined systems in 
real space revealed the splitting and merging of edge states from 
different bulk bands, which is also indicative of their topological nature. 

We also determined the wave functions of eigenstates in a completely trapped 
system and showed how these depend on the confining potential. With these, we 
calculated the dynamical structure factor which can, for 
instance, be measured by Bragg spectroscopy. We found that the dynamical 
structure factor can reveal the edge and bulk states of the system and their overlap. 

In summary, we demonstrated that topological properties in ultracold atomic systems with artificial gauge fields are not sensitive to
the trapping potentials available in optical lattice experiments
and that the edge states of these systems can be clearly detected via
Bragg spectroscopy. We believe that soft boundaries provide more detailed insight into the behavior of edge states, 
which cannot be observed in hard-wall systems, and are therefore worth investigation in their own right.\\
\begin{acknowledgments}We thank Ulf Bissbort, Peter Orth and Christiane de Morais-Smith for fruitful discussions. M.B. acknowledges support by the German National Academic Foundation and the Stiftung Polytechnische Gesellschaft Frankfurt am Main. This work was supported by the DFG via Sonderforschungsbereich SFB-TR/49 and Forschergruppe FOR~801. 
\end{acknowledgments}


\begin{thebibliography}{29}
\bibitem{CylinderGeometry}
\changed{The stripe geometry is infinitely extended in the $y$-direction and as long as translational invariance
is not broken along this direction, the Hamiltonian can be expressed in terms of $k_y$ and $x$. In this case, a
$100\times100$ lattice corresponds to $100$ lattice sites in the $x$-direction and $k_y$ being chosen on a 
$100$ points grid in the interval $[-\pi/a,\pi/a]$.}
\bibitem{QN}
Please note that the stripe geometry has the discrete translational invariance of an infinite lattice in $y$-direction and therefore the quasi-momentum $k_y$ is a well defined quantum number.
\bibitem{Remark5}
The integrated spectral function for the full system 
can be obtained from $\rho(k_y,\omega)=\rho_L(k_y,\omega)+\rho_R(k_y,\omega)$, where 
$\rho_R(k_y,\omega)$ is the integrated spectral function for the right half of the system,
which fulfills $\rho_R(k_y,\omega)=\rho_L(-k_y,\omega)$.
\bibitem{Disordered}
\changed{In the presence of a disordered background potential, translational invariance in the
$y$-direction is broken and therefore, to mimic a semi-infinite system, we diagonalize the Hamiltonian
of the system on a $L_x\times L_y$ lattice with periodic boundary conditions in the $y$-direction (i.e.
$\psi(x,y)=\psi(x,y+L_y)$, where $\psi(x,y)$ is the single-particle wave function).}
\bibitem{Remark3}
At this point, one has to keep in mind that we are investigating the properties of a finite system, which has per construction only localized eigenstates. Localization or delocalization therefore means confinement to few lattice sites or an extension over various sites and must be understood from the context.
\bibitem{Remark4}
In this case we used the Fermi velocity $v_F\equiv \partial_{k_y} 
\epsilon_F(k_y)$ of the edge states of the cylindrical geometry, where $k_y$ is 
a good quantum number, to approximate the frequency $\omega$, which is 
reasonable when looking at Fig.~\ref{fig:a16_compconf}.
\end{thebibliography}
\end{document}